\documentclass[conference,a4paper]{IEEEtran}

\usepackage{amsmath,amssymb,amsmath,amsfonts}
\usepackage{bm}
\usepackage{bbm}
\usepackage{graphicx}
\usepackage{cite}
\usepackage{epstopdf}
\usepackage{gensymb}
\usepackage{color}
\usepackage{lipsum}
\usepackage{float}
\usepackage{epstopdf}
\usepackage[mathcal]{eucal}
\usepackage{subfiles}
\usepackage[T1]{fontenc}
\usepackage{siunitx}
\usepackage[hyphens]{url}
\usepackage[breaklinks=true]{hyperref}
\usepackage{tabulary}
\usepackage{epsfig,graphics,graphicx,subfig,amssymb,amstext,amsmath,algorithm,algorithmic,wrapfig,multirow}
\usepackage{array}
\usepackage{adjustbox}
\usepackage[dvipsnames]{xcolor}
\usepackage{cite}
\usepackage{times}
\usepackage{placeins}
\usepackage{textcomp}
\usepackage[font=small]{caption}
\usepackage[breaklinks=true]{hyperref}
\usepackage{flushend}
\usepackage{color, colortbl}
\usepackage{tabularx}
\usepackage{bm}
\usepackage{enumerate}
\usepackage{tikz}
\usepackage{pgfplots}
\usepackage{epstopdf}
\usetikzlibrary{shapes,arrows}
\usepackage[utf8]{inputenc}
\usepackage{mathtools}
\usepackage[utf8]{inputenc}
\usepackage{xcolor}
\usepackage{tikz}
\usetikzlibrary{chains,arrows,calc,positioning}
\usepackage{amssymb}
\usepackage{pifont}
\usepackage{soul}
\usepackage{breqn} 
\usepackage{booktabs} 
\usepackage{multirow}
\usepackage{enumitem}
\usepackage{tabulary}
\usepackage[normalem]{ulem}
\usepackage{dblfloatfix}
\usepackage{makecell}
\usepackage{lipsum}
\usepackage{amsthm}
\usepackage{comment}
\usepackage{filecontents}

\newtheoremstyle{mystyle}
  {}
  {}
  {\itshape}
  {}
  {\bfseries}
  {.}
  { }
  {}
\theoremstyle{mystyle}

\def\ISDs   {{\mathsf{ISD}_{\sf s}}}

\newlength \figwidth
\definecolor{bittersweet}{rgb}{1.0, 0.44, 0.37}
\definecolor{glaucous}{rgb}{0.38, 0.51, 0.71}
\definecolor{gainsboro}{rgb}{0.86, 0.86, 0.86}
\definecolor{babyblueeyes}{rgb}{0.63, 0.79, 0.95}
\definecolor{silver}{rgb}{0.75, 0.75, 0.75}
\definecolor{neoncarrot}{rgb}{1.0, 0.64, 0.26}
\definecolor{Gray}{gray}{0.9}
\definecolor{LightCyan}{rgb}{0.88,1,1}
\definecolor{BackgroundLightBlue}{rgb}{0.97,0.97,1}
\definecolor{BackgroundGray}{gray}{0.98}

\makeatletter

 \let\oldforeign@language\foreign@language
 \DeclareRobustCommand{\foreign@language}[1]{%
   \lowercase{\oldforeign@language{#1}}}

\def\nb0{{\mathbf{0}}}
\def\nb1{{\mathbf{1}}}

\def\ncalR{{\mathcal{R}}}
\def\ncalS{{\mathcal{S}}}

\def\ncalU{{\mathcal{U}}}

\def\nrmx{{\rm x}}

\def\sinr{\mathtt{SINR}}			

\def\calN{\mathcal{N}}

\def\calT{\mathcal{T}}

\makeatother

\IEEEoverridecommandlockouts

\begin{document}

\bstctlcite{IEEEexample:BSTcontrol}

\title{UAV Communications in Integrated Terrestrial and Non-terrestrial Networks}

\author{
\IEEEauthorblockN{Mohamed Benzaghta$^{\star}$, Giovanni Geraci$^{\star}$, Rasoul Nikbakht$^{\flat}$, and David L\'{o}pez-P\'{e}rez$^{\sharp}$ \vspace{0.1cm}
}
\IEEEauthorblockA{$^{\star}$\emph{Universitat Pompeu Fabra (UPF), Barcelona, Spain}}
\IEEEauthorblockA{$^{\flat}$\emph{Centre Tecnologic de Telecomunicacions de Catalunya (CTTC), Barcelona, Spain}}
\IEEEauthorblockA{$^{\sharp}$\emph{Huawei Technologies, Boulogne-Billancourt, France}}
\thanks{M.~Benzaghta and G. Geraci were in part supported by MINECO's Project RTI2018-101040 and by a ``Ram\'{o}n y Cajal" Fellowship. Part of the work of R. Nikbakth was carried out while he was with Universitat Pompeu Fabra.}
}

\maketitle

\begin{abstract}
With growing interest in integrating terrestrial networks (TNs) and non-terrestrial networks (NTNs) to connect the unconnected, a key question is whether this new paradigm could also be opportunistically exploited to augment service in urban areas. We assess this possibility in the context of an integrated TN-NTN, comprising a ground cellular deployment paired with a Low Earth Orbit (LEO) satellite constellation, providing sub-6 GHz connectivity to an urban area populated by ground users (GUEs) and uncrewed aerial vehicles (UAVs). Our study reveals that offloading UAV traffic to the NTN segment drastically reduces the downlink outage of UAVs from 70\% to nearly zero, also boosting their uplink signal quality as long as the LEO satellite constellation is sufficiently dense to guarantee a minimum elevation angle. Offloading UAVs to the NTN also benefits coexisting GUEs, preventing uplink outages of around 12\% that GUEs would otherwise incur. Despite the limited bandwidth available below 6\,GHz, NTN-offloaded UAVs meet command and control rate requirements even across an area the size of Barcelona with as many as one active UAV per cell. Smaller UAV populations yield proportionally higher rates, 
potentially enabling aerial broadband applications.
\end{abstract}

\section{Introduction}

Thanks to their low cost and high mobility, 
uncrewed aerial vehicles (UAVs) may soon take over important tasks including search and rescue, delivery, and remote sensing. 
In the next decade, 
UAV taxis may also redefine how we commute and, in turn, where we live and work. 
For these and other applications, 
UAVs will transfer real-time data to and from the mobile network, 
requiring reliable connectivity for command and control (C\&C) and mission-specific data payloads~\cite{geraci2021will,zeng2019accessing,wu20205g,MozSaaBen2018,BucBerBas2022,CheJaaYan2020,3GPP22.125}.

Due to their height, however,
UAVs receive/create line-of-sight (LoS) interfering signals from/to a plurality of cells, 
respectively hindering the decoding of C\&C messages and disrupting the service of legacy ground users (GUEs)~\cite{ZenGuvZha2020,SaaBenMoz2020,AbdPowMar2020}.
Technical solutions to this problem have been introduced in 4G LTE~\cite{3GPP36777} and 5G NR~\cite{3GPPRP212715} to deal with a handful of connected UAVs. 
More advanced solutions hinge, e.g., on uptilted cells~\cite{chowdhury2021ensuring,XiaRanMez2020a,KanMezLoz2021}, massive MIMO~\cite{GarGerLop2019}, or cell-free architectures~\cite{DanGarGer2020}, 
and may be less economically viable in the short term, 
as they necessitate dedicated network upgrades. 

As an alternative to ground densification, 
UAVs could be supported by non-terrestrial networks (NTNs), 
e.g., via Low Earth Orbit (LEO) satellite constellations~\cite{SedFelLin2020,RinMaaTor2020,kodheli2020satellite,giordani2020non}. 
While mainly targeting underserved areas, NTNs may also be leveraged to augment urban connectivity, e.g., with a terrestrial network (TN) operator opportunistically leasing spectrum and infrastructure from an NTN one. 
Indeed, the ongoing 3GPP efforts towards TN-NTN integration will allow mobile devices to seamlessly switch from one segment to the other~\cite{GerLopBen2022,3GPP38821,3GPP38811}.

In this paper, 
we consider an integrated TN-NTN, 
comprising a ground cellular deployment paired with a LEO satellite constellation, 
providing sub-6 GHz connectivity to an urban area populated by ground users (GUEs) and UAVs. 
We study the benefits of offloading the latter to the NTN, 
both in the downlink (DL) and uplink (UL), 
for different satellite elevation angles and beam reuse schemes, 
while also accounting for the main propagation features, antenna models, and deployment scenarios specified by the 3GPP. 
Our main findings can be summarized as follows:
\begin{itemize}
\item 
UAVs connected to a standalone TN incur a nearly 70\% downlink outage when flying at 150m due to ground-to-air interference. 
Offloading UAV traffic to the NTN segment ensures reliable coverage, 
also outperforming a TN that implements interference coordination.
\item 
TN-connected UAVs generate strong uplink interference, 
forcing---in the most challenging scenario---up to 12\% of the GUEs into outage, 
even under practical fractional power control. 
Such outage is reduced to below 1\% when UAV uplink traffic is offloaded to the NTN.
\item 
Under inter-beam interference, 
NTN-offloaded UAVs may experience outage when the NTN elevation angle falls to $87^{\circ}$ and below, 
calling for a careful design of beam reuse according to the LEO constellation density.
\item 
For an area the size of Barcelona, 
the NTN can provide C\&C rates of 60-100~kbps for as many as one UAV per TN cell. 
Lower UAV penetration yields proportionally higher rates, 
potentially enabling broadband use-cases.
\end{itemize}

\section{System Model}

In this section, we introduce the network topology and channel models used. 
Further details are given in Table~\ref{table:parameters}.

\begin{figure}
\centering
\includegraphics[width=\figwidth]{
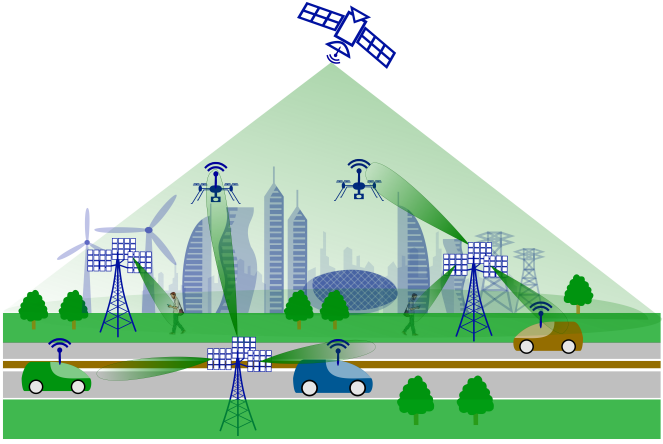}
\caption{Illustration of an integrated TN-NTN comprising a terrestrial and a satellite segment serving GUEs and UAVs in an urban area.}
\label{fig:deployment}
\vspace{-0.6cm}
\end{figure}


\subsection{Integrated TN-NTN Deployment}

We consider a cellular TN as specified by 3GPP~\cite{3GPP38901,3GPP36777}. 
We also assume the availability of an NTN segment through a LEO satellite orbiting at 600~km, whose features are chosen according to~\cite{3GPP38821,3GPP38811}.
\footnote{As the penetration of UAVs increases, 
a TN operator intending to provide aerial connectivity may choose to lease infrastructure and spectrum from an NTN operator \cite{GerLopBen2022}. While the latter would primarily deploy and operate LEO constellations to cover currently underserved areas, 
their inherent mobility makes them available over urban areas as well, 
generating opportunities in the multi-operator scenario considered in this paper.}
We consider handheld GUEs and UAVs, all capable of connecting to either the TN or NTN.

\subsubsection*{Terrestrial network}

Base stations (BSs) are deployed on a hexagonal layout, 
and communicate with their respective sets of connected users in downlink and uplink. 
Deployment sites are comprised of three co-located BSs, 
each covering one sector---i.e., a TN cell---spanning an angular interval of $120^{\circ}$. 
For an inter-site distance $\mathrm{ISD_{TN}}$, 
the area of a TN cell is given by $A_{\mathrm{TN}} = \sqrt{3} \cdot \mathrm{ISD_{TN}}^2 / 6$.

\subsubsection*{Non-terrestrial network}
 
The LEO satellite segment generates multiple beams to serve users in both downlink and uplink. 
The beams point to the ground in a hexagonal fashion, 
each creating one corresponding NTN cell. 
The spacing between adjacent beams is computed according to the half-power beamwidth (HPBW) of each beam's radiation pattern.
Due to its orbital movement, the LEO satellite is seen under different elevation angles from the standpoint of the TN, and accordingly, the footprint of its Earth-moving beams illuminates different areas on the ground. 
We study how the NTN performance is affected by the elevation angle, since this has implications on the LEO constellation density needed.

\subsubsection*{User population}

Both the TN and NTN are capable of serving GUEs and UAVs, 
e.g., providing the former with data and the latter with data and C\&C information. 
We assume the total user population to be concentrated in an urban area of size $A_{\mathrm{U}}$, contained in an NTN cell and 
resulting in a total of $A_{\mathrm{U}} / A_{\mathrm{TN}}$ TN cells.
In this area, 
GUEs are located both outdoor, at height $h_{\rm out}=1.5$~m, and indoor, at $h_{\rm in}$ in buildings consisting of several floors. UAVs fly outdoor at height $h_{\rm UAV}$.

\subsubsection*{Spectrum allocation}

We assume the TN and NTN to employ orthogonal bands and frequency division duplexing (FDD). 
For both the TN and NTN, 
we assume the nominal downlink and uplink carrier frequencies to be at 2~GHz \cite{3GPP38821}. 
For the TN, 
the available bandwidth is fully reused across all cells. 
For the NTN, 
we consider two possible frequency reuse factors (FRF), 
namely: (i) FRF = 1, where all frequency resources are fully reused across all beams, 
and (ii) FRF = 3, 
where they are partitioned into three sets, 
each reused every three beams. 
As it will be shown, 
the latter approach sacrifices peak performance in favor of reducing interference between adjacent beams and in turn enhance cell-edge performance.

\subsection{Propagation Channel}

The main propagation channel features are described in the following and summarized in Table~\ref{table:parameters}.

\subsubsection*{Path loss and shadow fading}

All radio links are affected by path loss and lognormal shadow fading, 
both dependent on the link LoS condition,
and modeled as in~\cite{3GPP36777,3GPP38811}. 
TN links explicitly account for the transmitter and receiver heights, 
as this is crucial when modeling cellular-connected UAV users, 
with higher UAVs more likely to experience LoS condition from a plurality of cells~\cite{3GPP36777}. 
NTN links account for the LEO satellite elevation angle, 
i.e., the angle between the line pointing towards the satellite and the local horizontal plane. 
Elevation angles closer to nadir, i.e., $90^{\circ}$, yield shorter LEO-to-user distances and are more likely to be in LoS~\cite{3GPP38811}. 
Compared to a TN link, 
the signal travelling on an NTN link undergoes several extra stages of propagation. 
As a result, 
the total path loss consists of additional terms accounting for the attenuation due to atmospheric gases and to scintillation.

\subsubsection*{Antenna gain}

We assume all GUEs and UAVs to have a single omnidirectional antenna with unitary gain. 
We assume each TN BS to be equipped with a vertical, downtilted, uniform linear array, with semi-directive elements and a single radio-frequency chain. 
The latter yields a realistic radiation pattern, modeling the upper sidelobes seen by UAVs \cite{AzaGerGar2020}. 
Lastly, we assume the antenna generating each LEO satellite beam to be a typical reflector with circular aperture \cite{3GPP38811}.

The total large-scale power gain on a link comprises path loss, shadow fading, and antenna gain at both transmitter and receiver. 
We denote $G_{\nrmx,k}$ the large-scale power gain between cell $\nrmx$ and user $k$, 
with the subscript $\nrmx \in \{t,n\}$ referring either to a specific TN cell $t$ or a specific NTN cell $n$.

\subsubsection*{Small-scale fading}

Similarly, 
we denote $h_{\nrmx,k}$ the small-scale block fading between cell $\nrmx$ and user $k$. 
We assume TN-connected GUEs to undergo Rayleigh fading and UAV links to experience pure LoS propagation conditions, 
given their elevated position with respect to the clutter of buildings.
\footnote{Different small-scale fading assumptions only yield anecdotal quantitative changes in the numerical results presented. 
They could however play a more prominent role when considering spatial multiplexing and antenna correlation.}

\subsubsection*{Power control (TN)}

In the downlink, 
we assume all TN BSs to transmit the same power $P_t$. 
In the uplink, 
we consider open-loop fractional power control for all users connected to the TN, 
as per the cellular systems currently deployed. 
Accordingly, the power $P_{k}$ transmitted by a given user $k$ is adjusted depending on the serving TN cell $t$ as~\cite{BarGalGar2018GC}
\begin{equation}
P_{k} = \min\left\{ P^{\textrm{max}}, P_0 \cdot G_{t,k}^{\alpha} \right\},
\label{eqn:power_control}
\end{equation}
where $P^{\textrm{max}}$ is the user's maximum transmit power, 
$P_0$ is a parameter adjusted by the network, 
the exponent $\alpha \in [0,1]$ is the fractional power control factor, 
and $G_{t,k}$ is the large-scale fading between user $k$ and TN cell $t$. 
The aim of (\ref{eqn:power_control}) is to compensate for a fraction $\alpha$ of the large-scale fading, 
up to a limit imposed by $P^{\textrm{max}}$. 

\subsubsection*{Power control (NTN)}

Uplink fractional power control is not applied to users connected to the NTN, 
with the latter transmitting at maximum power $P_k=P^{\textrm{max}}$. 
Similarly, a fixed downlink power $P_n$ is always used by each NTN cell \cite{3GPP38821}. 

\begin{table}
\centering
\caption{System parameters}
\label{table:parameters}
\def\arraystretch{1.2}
\begin{tabulary}{\columnwidth}{ |p{2.1cm} | p{5.85cm} | }
\hline
	\textbf{TN deployment} 			&  \\ \hline
  Cell layout				& Hexagonal grid over an area $A_{\mathrm{U}} = \SI{52}{km^2}$, $\mathrm{ISD_{TN}} = 500$~m, three sectors per site, one BS per sector at $25$~m   \cite{3GPP36777} \\ \hline
  Frequency band 		& FRF=1, $B_{\nrmx} \!=\!$ 10+10 MHz (DL+UL) at 2~GHz \cite{3GPP36777} \\ \hline
	BS transmit power 			& 46~dBm \cite{3GPP36777} \\ \hline   
	Antenna elements 		& Horiz./vert. HPBW: $65^{\circ}$, max. gain: 8~dBi \cite{3GPP36777} \\ \hline
	Antenna array 		& $10\times 1$, downtilt: $12^{\circ}$, element spacing: $0.5\lambda$ \cite{3GPP36777}\\ \hline
	Noise figure 			& 7~dB \cite{3GPP38901} \\ \hline \hline
	
	\textbf{NTN deployment} 			&  \\ \hline
  Cell layout				& Orbit: 600~km, 7 beams centered on a hexagonal grid, elevation angle: variable \cite{3GPP38821} \\ \hline
  \multirow{2}{*}{Frequency band}		& FRF=1, $B_{\nrmx}\!=\!$ 30+30 MHz (DL+UL) at 2~GHz \cite{3GPP38821}  \\ \cline{2-2} 			& FRF=3, $B_{\nrmx}\!=\!$ 10+10 MHz (DL+UL) at 2~GHz \cite{3GPP38821} \\ \hline
			Transmit power			& 34 dBW/MHz per beam \cite{3GPP38821} \\ \hline
  Beam antenna				& Reflector with circular aperture, HPBW: 4.41$^{\circ}$, max. gain: 30~dBi \cite{3GPP38821} \\ \hline
  $G/T$			& 1.1 dB/K antenna gain-to-noise-temperature \cite{3GPP38821} \\ \hline\hline

	\textbf{Users} 			&  \\ \hline
  User distribution 				& 15 users per TN sector on average \cite{3GPP36777} \\ \hline
	\multirow{2}{*}{GUE distribution} 				& 80\% indoor, horiz.: uniform, $h_{\rm in}$: uniform in buildings of four to eight floors  \cite{3GPP38901} \\ \cline{2-2}
	 				& 20\% outdoor, horiz.: uniform, $h_{\rm out}=1.5$~m \cite{3GPP38901} \\ \hline
	UAV distribution 				& Outdoor, horiz.: uniform, $h_{\rm UAV}$: 150~m \cite{3GPP36777} \\ \hline
	UAVs/GUEs ratio 				& 3GPP Case~2: 0.7\%, Case~3: 7.1\% \cite{3GPP36777} \\ \hline
	User association				& Based on RSRP (large-scale fading) \\ \hline
	\multirow{2}{*}{Scheduler}		& DL: single-user round robin, $B_k=B_{\nrmx}$  \\ \cline{2-2} 
			& UL: multi-user round robin, $B_k=360$~kHz \cite{SedFelLin2020} \\ \hline
		\multirow{2}{*}{UL power control}		& TN: fractional power control with $\alpha = 0.80$, $P_{0} = -85$~dBm, and $P_{\textrm{max}}=23$~dBm \cite{3GPP36777} \\ \cline{2-2} 
			& NTN: always max power $P_{\textrm{max}}=23$~dBm \cite{3GPP38821}\\ \hline
	User antenna 		& Omnidirectional, gain: 0~dBi \cite{3GPP36777} \\ \hline
	Noise figure 			& 9~dB \cite{3GPP36777} \\ \hline \hline

	\textbf{Channel model} 			&  \\ \hline
	Large-scale fading 		& Urban Macro as per \cite{3GPP36777, 3GPP38901,3GPP38811} \\ \hline
	\multirow{2}{*}{Small-scale fading}		& TN-GUE: Rayleigh  \\ \cline{2-2} 
			& TN-UAV and NTN-UAV: pure LoS \\ \hline
	Thermal noise 				& -174 dBm/Hz spectral density \cite{3GPP36777}\\ \hline 
\end{tabulary}
\vspace{-0.5cm}
\end{table}

\section{Evaluation Methodology}

To study the gains provided by a NTN segment, 
we compare the performance of GUEs and UAVs in an integrated TN-NTN to the one experienced in a baseline standalone TN. 
In this section, 
we introduce the association and user offloading policy adopted in each case. 
Furthermore, 
we elaborate on the key performance indicators of interest.

\subsection{Cell Association and User Offloading}

We denote as $\calT$ and $\calN$ the sets of TN and NTN cells, respectively, and as $\ncalU_{\nrmx}$ the set of active users associated to cell $\nrmx$. We denote $\ncalS_{\rm T}$ and $\ncalS_{\rm N}$ as the set of users served by the TN and NTN, respectively.

\subsubsection*{Standalone TN}

In the conventional case of a standalone TN, 
all GUEs and UAVs associate to the TN cell providing the largest reference signal received power (RSRP). 
For user $k$, 
the latter corresponds to the cell $\nrmx$ providing the highest large-scale gain $G_{\nrmx,k}$.

\subsubsection*{UAV offloading to the NTN}

In the case of an integrated TN-NTN, 
we assume all GUEs to be served by the TN and UAV users to be offloaded to the NTN segment.
\footnote{The study of a case-by-case offloading policy, 
accounting for factors such as the UAV height and the cell load will be the subject of future work.} 
As it will be clear in the numerical results section, 
this choice is triggered by the mutual interference between aerial and ground transmissions, where the performance of UAVs (resp. GUEs) can be severely impaired in the downlink (resp. uplink)~\cite{GarGerLop2019}.

\subsection{Key Performance Indicators}

Under the orthogonal spectrum allocation between TN and NTN, the DL and UL signal-to-interference-plus-noise ratios (SINRs) on a given time-frequency physical resource block (PRB) of user $k$ served by TN cell $t$ are respectively given by
\begin{equation}
  \vspace{-0.2cm}
  \sinr_{t,k}^{\text{DL}} = \frac{P_t \cdot G_{t,k} \cdot |h_{t,k}|^2 }{
  \sum\limits_{\tau\in\calT\backslash t}{P_{\tau} \cdot G_{\tau,k} \cdot |h_{\tau,k}|^2 + \sigma^2_k}},
  \label{SINR_DL_TN}
  \vspace{-0.1cm}
\end{equation}
\begin{equation}
  \sinr_{t,k}^{\text{UL}} = \frac{P_k \cdot G_{t,k} \cdot |h_{t,k}|^2}{
  \sum\limits_{\ell\in\ncalS_{\rm T}\backslash k}{P_{\ell} \cdot G_{t,\ell} \cdot |h_{t,\ell}|^2 + \sigma^2_k}},
  \label{SINR_UL_TN}
\end{equation}
where $\sigma^2_k$ is the thermal noise variance over the bandwidth $B_k$ accessed by user $k$. Similarly, 
the DL and UL SINRs per PRB of user $k$ served by NTN cell $n$ are respectively given by
\begin{equation}
  \vspace{-0.2cm}
  {\sinr_{n,k}^{\text{DL}} =  \frac{P_n \cdot G_{n,k} \cdot |h_{n,k}|^2 }{
  \sum\limits_{\nu\in\calN\backslash n}{P_{\nu} \cdot G_{\nu,k} \cdot |h_{\nu,k}|^2 + \sigma^2_k}},}
  \label{SINR_DL_NTN}
    \vspace{-0.1cm}
\end{equation}
\begin{equation}
  {\sinr_{n,k}^{\text{UL}} = \frac{P_k \cdot G_{n,k} \cdot |h_{n,k}|^2}{
  \sum\limits_{\ell\in\ncalS_{\rm N}\backslash k}
  {P_{\ell} \cdot G_{n,\ell} \cdot |h_{n,\ell}|^2 + \sigma^2_k}}.}
  \label{SINR_UL_NTN}
\end{equation}

The rate $\ncalR_k$ achievable by user $k$ served by cell $\nrmx$ can be related to its SINR (either in downlink or uplink) as
\begin{equation}
    \ncalR_k = \eta_k B_k \, \mathbb{E} \left[ \log_2 (1 + \sinr_{\nrmx,k}) \right],
    \label{rates}
\end{equation}
with $\nrmx$ denoting the serving TN or NTN cell depending on the user's association, $B_k$ the bandwidth allocated to user $k$, and $\eta_k$ the fraction of time user $k$ is scheduled by cell $\nrmx$, and where the expectation is taken over the small-scale fading.
In the downlink, 
we assume each cell $\nrmx$ to multiplex its set of associated users $\ncalU_{\nrmx}$ in the time domain, 
yielding $\eta_k = \vert \ncalU_{\nrmx} \vert^{-1}$, 
and to allocate the entire available band $B_{\nrmx}$ to the scheduled user $k$, i.e., $B_k = B_{\nrmx}$.
In the uplink, we assume user multiplexing to occur both in the time and frequency domains, with each user allocated a bandwidth $B_k$ and $N_{\nrmx} = \min \{B_{\nrmx}/B_k, \vert \ncalU_{\nrmx} \vert\}$ users scheduled simultaneously, each for a fraction of time $\eta_k = N_{\nrmx} / \vert \ncalU_{\nrmx} \vert$. 
This choice is owed to the limited power budget available at the user~\cite{3GPP38821}.
\section{Numerical Results}

In this section, 
we conduct downlink and uplink experiments to evaluate the performance gains provided by UAV traffic offloading in an integrated TN-NTN. 
We consider UAVs flying at $h_{\rm UAV}=$150~m, 
which is known to be a challenging scenario for GUE-UAV coexistence~\cite{GerGarGal2018}.

\subsection{Downlink Experiments}

\begin{figure}
\centering
\includegraphics[width=\figwidth]{
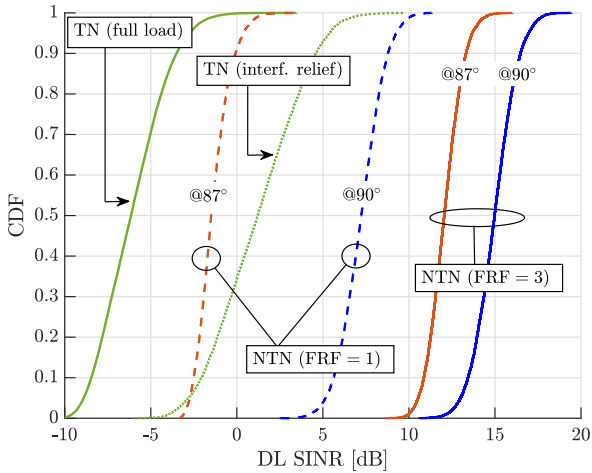}
\caption{CDF of the downlink SINR per PRB experienced by UAVs when connected to a standalone TN and when offloaded to a NTN segment. For the latter, two LEO satellite elevation angles and two beam frequency reuse patterns are considered. The SINR experienced from a TN with interference relief is also shown for comparison.}
\label{fig:SINR_DL}
\vspace{-0.5cm}
\end{figure}

\begin{figure*}
\centering
\includegraphics[width=0.92\textwidth]{
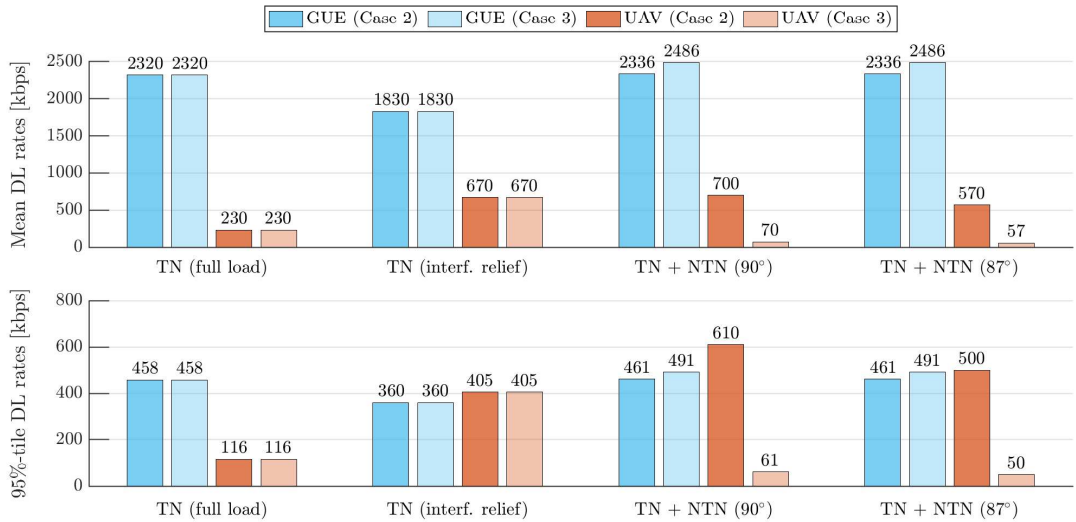}
\caption{Mean and $95\%$-tile downlink rates for GUEs and UAVs when: (i) all are served by a fully loaded TN; (ii) interference relief is guaranteed at UAVs through reserved resources; (iii) and (iv) UAVs are offloaded to a NTN segment at $90^{\circ}$ and $87^{\circ}$, respectively, with FRF = 3. Both 3GPP Case 2 and Case 3 are considered, respectively corresponding to a penetration of 72 and 720 UAVs in the urban area.}
\label{fig:Capacity_DL}
\vspace{-0.5cm}
\end{figure*}

Fig.~\ref{fig:SINR_DL} shows the CDF of the downlink SINR per PRB experienced by UAVs when connected to a standalone TN and when offloaded to a NTN segment. 
For the latter, 
two LEO satellite elevation angles ($90^{\circ}$ and $87^{\circ}$) and two beam frequency reuse patterns (FRF = 1 and FRF = 3) are considered.\footnote{Note that the downlink SINR does not depend on the UAV penetration.} 
For comparison, 
the figure also shows the SINR experienced by UAVs from a TN when the dominant interfering TN cells are turned off to guarantee a minimum SINR of -5~dB (a proxy for coverage \cite{GerGarGal2018}). 
The following observations can be made:
\begin{itemize}[leftmargin=*]
    \item 
    Standalone TNs struggle to provide reliable coverage to high UAVs, 
    with their SINR falling below -5~dB in 70\% of the cases (solid green). 
    This is in line with previous results~\cite{geraci2021will}.
    \item 
    Offloading UAVs to the NTN segment ensures coverage, 
    yielding SINRs ranging between -3~dB and 17~dB depending on the LEO satellite elevation angle and the beam frequency reuse pattern considered, as discussed in what follows.
    \item 
    Moving from FRF = 3 to FRF = 1 entails full reuse and thus inter-beam interference. 
    The latter degrades the median SINR by approximately 8~dB (solid vs. dashed blue) and 14~dB (solid vs. dashed red) for a LEO satellite at elevation angles of $90^{\circ}$ and $87^{\circ}$, respectively.
    \item 
    The UAV SINR experiences a prominent degradation when the LEO satellite moves from $90^{\circ}$ to $87^{\circ}$, 
    due to a larger propagation distance and a lower antenna gain. 
    Indeed, given the HPBW of $4.41^{\circ}$, 
    an elevation angle of $87^{\circ}$ implies being served nearly at the edge of a beam. 
    Nonetheless, all NTN-offloaded UAVs still remain in coverage, 
    even in the presence of inter-beam interference (dashed red).
    \item 
    By design, TN interference relief through cell switch-off guarantees coverage for all UAVs (dotted green). 
    However, the resulting SINRs are lower than those attainable via NTN offloading. 
    Moreover, such approach sacrifices a considerable amount of radio resources, 
    since relieving each UAV of the dominant interference requires approximately 12 TN cells to be idle. 
    The latter results in a capacity loss for legacy GUEs, 
    quantified in the sequel.
\end{itemize}

Fig.~\ref{fig:Capacity_DL} shows the mean and $95\%$-tile downlink rates achievable by GUEs and UAVs in 3GPP Cases 2 and 3, 
respectively corresponding to one active UAV every 10 TN cells and one active UAV per TN cell. 
In this figure, 
FRF = 3 is assumed, 
corresponding to higher SINRs on the NTN segment, 
but also to a reduction of 2/3 of the available bandwidth. 
The rates computation follows the methodology in (\ref{rates}), 
assuming users distributed over an urban area $A_{\mathrm{U}} = \SI{52}{km^2}$, 
roughly the size of the city of Barcelona. 
Fig.~\ref{fig:Capacity_DL} considers four different scenarios: 
(i) a standalone TN serving both GUEs and UAVs; 
(ii) a standalone TN where dominant interfering cells are switched off on certain PRBs to ensure UAV coverage; 
(iii) and (iv) an integrated TN-NTN where UAV traffic is offloaded to a LEO satellite seen at $90^{\circ}$ and $87^{\circ}$, respectively.
Fig.~\ref{fig:Capacity_DL} carries the following messages:
\begin{itemize}[leftmargin=*]
    \item 
    In Case 2---corresponding to 72 active UAVs over the urban area considered---offloading downlink UAV traffic to the NTN segment results in a remarkable five-fold increase in the UAV rates when the LEO satellite is at $90^{\circ}$.
    \item 
    Owing to a more stable NTN link budget, 
    the rates of NTN-offloaded UAVs exhibit less variance than their TN counterpart, 
    with a small relative gap between mean and $95\%$-tile rates. 
    In all scenarios considered, 
    the mean UAV rates remain around the recommended 60--100~kbps for C\&C \cite{3GPP36777}. 
    This occurs even in Case 3---corresponding to 720 active UAVs over the urban area---and when the LEO satellite moves from $90^{\circ}$ to $87^{\circ}$. 
    \item 
    In a standalone TN, 
    DL rates are independent of the UAV penetration assumed, 
    since the total number of users served by each cell remains unchanged. 
    However, offloading UAVs to the NTN slightly increases the GUE rates, as the capacity of each TN cell has to be shared among fewer users.
    \item 
    Standalone TNs with interference relief guarantee similar mean rates for the UAVs as the NTN does in Case 2. 
    This however comes at the expense of reducing the legacy GUE rates, 
    a shortcoming that NTN offloading does not entail.
\end{itemize}

\subsection{Uplink Experiments}

\begin{figure}
\centering
\includegraphics[width=\figwidth]{
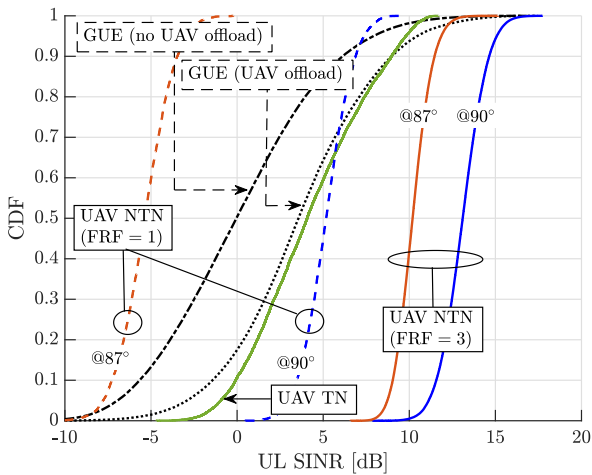}
\caption{CDF of the uplink SINR per PRB experienced by GUEs and UAVs in Case 3 when all are connected to a standalone TN and when UAVs are offloaded to a NTN segment. For the latter, two LEO satellite elevation angles and two FRF values are considered.}
\label{fig:SINR_UL}
\vspace{-0.5cm}
\end{figure}

\begin{figure*}
\centering
\includegraphics[width=0.92\textwidth]{
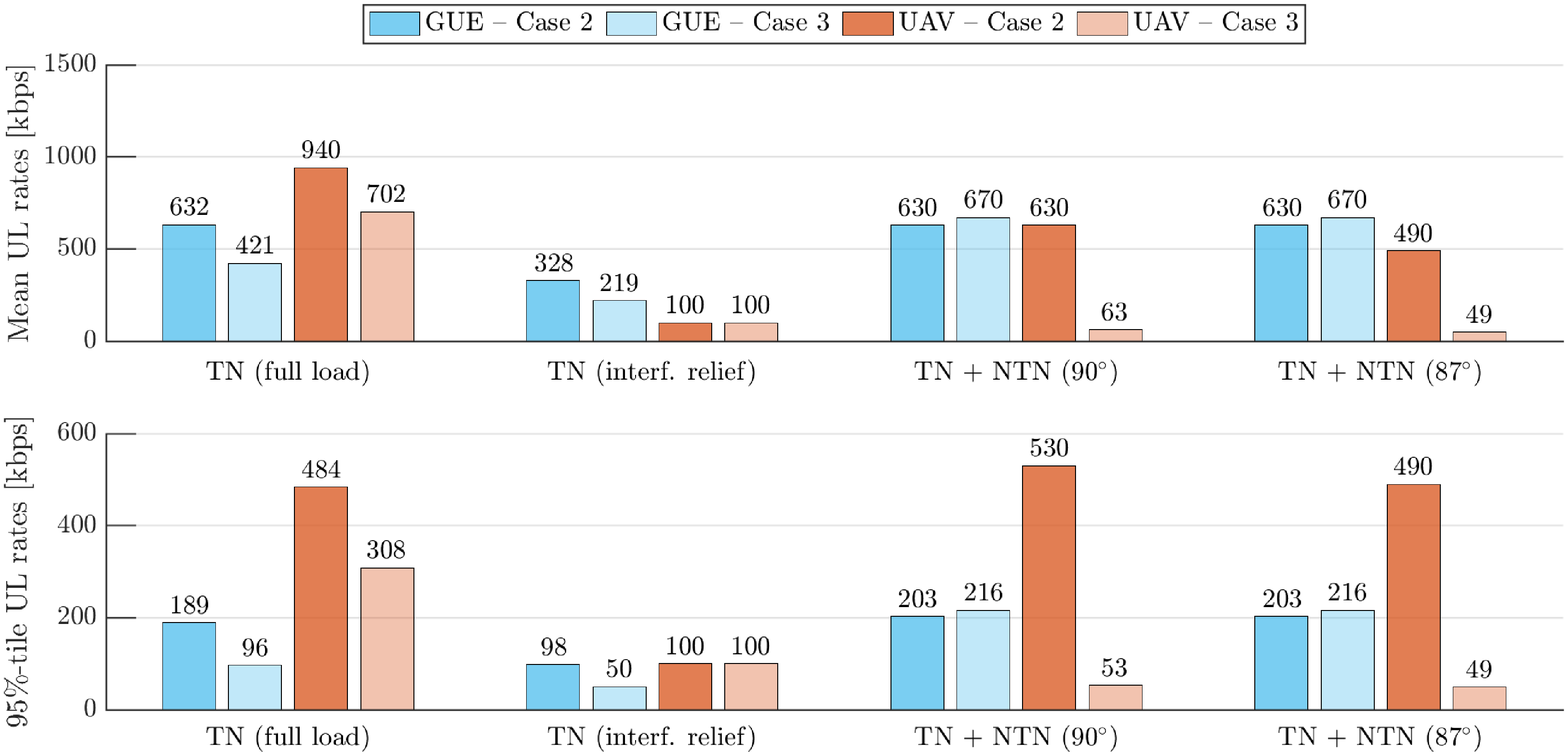}
\caption{Mean uplink rates for GUEs and UAVs when: (i) all are served by a fully loaded TN; (ii) a standalone TN reserves radio resources to UAVs to guarantee 100~kbps C\&C rates; (iii) UAVs are offloaded to a NTN segment at $90^{\circ}$; and (iv) UAVs are offloaded to a NTN segment at $87^{\circ}$. 3GPP Case 2 and Case 3 are both considered for the UAV density.}
\label{fig:Capacity_UL}
\vspace{-0.5cm}
\end{figure*}

Fig.~\ref{fig:SINR_UL} shows the CDF of the uplink SINR per PRB experienced by GUEs and UAVs in Case 3, 
when all are connected to a standalone TN and when UAV traffic is offloaded to a NTN segment. 
For the latter, 
two LEO satellite elevation angles and two beam frequency reuse patterns are considered. 
The following remarks can be made:
\begin{itemize}[leftmargin=*]
    \item 
    Even in the presence of fractional power control, 
    the uplink interference generated by UAVs may jeopardize the GUE performance, 
    causing their SINR to drop below -5~dB in 12\% of the cases (dash-dotted black).
    \footnote{Assigning and tuning separate power control parameters for GUEs and UAVs may reduce the GUE outage at the expense of the UAV performance.}
    Such outage is drastically reduced to only 1\% by offloading the UAV traffic to the NTN (dotted black).
    \item 
    When offloaded to the NTN, 
    UAVs see a boost in their uplink SINR as long as the inter-beam interference is kept at bay (solid green vs. blue/red curves). 
    While the median SINR gain can reach up to 9~dB over the TN baseline in ideal conditions (LEO satellite at $90^{\circ}$ and FRF = 3, solid blue), 
    some UAVs may be forced into outage in worse scenarios (LEO satellite at $87^{\circ}$ with full frequency reuse, dashed red).
\end{itemize}

Fig.~\ref{fig:Capacity_UL} completes the picture by showing the mean and $95\%$-tile uplink rates for GUEs and UAVs when: 
(i) all are served by a fully loaded TN; 
(ii) the TN employs interference coordination, 
allocating GUEs and UAVs separate radio resources in a proportion designed to guarantee 100~kbps C\&C rates at the UAVs; 
(iii) and (iv) UAVs are offloaded to a NTN segment at $90^{\circ}$ and at $87^{\circ}$, respectively. 
Again, 3GPP Case 2 and Case 3 are both considered for the penetration of UAV users. 
Fig.~\ref{fig:Capacity_UL} prompts the following important observations:
\begin{itemize}[leftmargin=*]
    \item 
    In Case 2, a standalone TN is capable of ensuring coverage to most GUEs and UAVs. 
    In Case 3, offloading uplink UAV traffic to the NTN enhances the GUE rates by about 50\%, 
    though the latter comes at the expense of reduced UAV rates.
    \item 
    Even in Case 3---with as many as one active UAV per cell across an urban area the size of Barcelona---and with a bandwidth of only 10~MHz per beam, NTN-offloaded UAVs achieve rates in the order of 60~kbps for C\&C. 
    \item 
    NTN-offloading outperforms a standalone TN that enforces time-frequency orthogonality between GUEs and UAVs, since the latter sacrifices a large amount of radio resources, resulting in poor data rates.
\end{itemize}
\section{Conclusion}

In this paper, we studied supporting UAV communications through an integrated TN-NTN operating below 6~GHz. By offloading UAV traffic to a LEO spaceborne segment, we aimed at guaranteeing reliable UAV C\&C coverage without degrading the performance of coexisting GUEs.
Our study revealed that NTN-offloaded UAVs flying at 150~m see their downlink outage drastically reduced from 70\% to nearly zero. UAVs also experience an uplink SINR boost as long as the LEO satellite constellation is sufficiently dense to guarantee a minimum elevation angle. Offloading UAVs to the NTN may also benefit legacy GUEs, particularly in the uplink as it prevents outages of around 12\% that GUEs would otherwise incur. 
Despite the limited bandwidth available below 6 GHz, NTN-offloaded UAVs meet minimum C\&C rate requirements even across an urban area the size of Barcelona with as many as one active UAV per cell. With the lower UAV penetration envisioned in the short-to-medium term, proportionally higher rates can be expected, enabling aerial broadband applications.

\bibliographystyle{IEEEtran}
\bibliography{journalAbbreviations,main}

\end{document}